\documentclass[12pt]{article}
\pdfoutput=1
\usepackage{amssymb}
\usepackage{amsmath}
\usepackage{amsfonts}
\usepackage{graphicx}
\usepackage{mathrsfs}
\usepackage{comment}
\usepackage{cite}
\usepackage{bm,color} 
\usepackage{amssymb,amsfonts,slashed,amsthm,amsmath,graphicx, soul}
\usepackage{epsfig}

\newcommand{\Slash}[1]{{\ooalign{\hfil#1\hfil\crcr\raise.167ex\hbox{/}}}}

\newcommand{\beq}{\begin{equation}}  \newcommand{\eeq}{\end{equation}}
\newcommand{\bef}{\begin{figure}}  \newcommand{\eef}{\end{figure}}
\newcommand{\bec}{\begin{center}}  \newcommand{\eec}{\end{center}}
  
\newcommand{\laq}[1]{\label{eq:#1}}

\def\({\left(}
\def\){\right)}

\def\O{\mathcal{O}}

\newcommand{\GEV}{ {\rm \, GeV} }

\def\a{\alpha}

\def\f{\phi}

\def\r{\rho}
\def\s{\sigma}

\def\L{\Lambda}
\def\F{\Phi}

\def\*{\dagger}

\begin{document}
\begin{center}
{\Large\bf Cosmic Clues: DESI, Dark Energy, and the Cosmological Constant Problem}  
\vspace{1.5cm}

{\bf Wen Yin}

\vspace{12pt}
\vspace{1.5cm}
{\em 

{Department of Physics, Tokyo Metropolitan University, Tokyo, Japan } }

\vspace{1.5cm}

\abstract{
Several attempts to solve the cosmological constant problem, which concerns the value of the cosmological constant being extremely smaller than the Standard Model mass scales, have introduced a scalar field with a very flat potential that can be approximated as linear around any given position.
The scalar field scans the cosmological constant in such a way that the current small value is explained. 
Recently, Dark Energy Spectroscopic Instrument (DESI) reported the results of the first year. Combining the data with CMB, Pantheon, Union3, and/or DES-SN5YR, there is a
  preference or anomaly, indicating that the dark energy in the current Universe slightly deviates from that in the $\Lambda$CDM model and varies over time. In this paper, I show that the simple linear potential of a scalar field that may explain the small cosmological constant, can explain the DESI anomaly. 
In particular, the model proposed by the present author in \cite{Yin:2021uus}, which relaxes the cosmological constant by the condition that inflation ends, predicts a time-dependence of the dark energy close to the one favored by the data.}

\end{center}
\clearpage

\setcounter{page}{1}
\setcounter{footnote}{0}

\setcounter{footnote}{0}

\section{Introduction} 

The Dark Energy Spectroscopic Instrument (DESI) collaboration has recently reported its first cosmological result from the precise baryon acoustic oscillation measurement in galaxy, quasar and Lyman-$\a$ forest tracers with a year's observation~\cite{DESI:2024mwx}. 
The paper reveals that allowing for time-variance of the dark energy at a late epoch, with or without curvature around the redshift $z=\mathcal{O}(0.1)$, yields a better fit than the $\Lambda$CDM either based on the combination of DESI data with cosmic microwave background (CMB) data or type Ia supernovae.\footnote{In inflationary cosmology, curvature is likely to be suppressed. I consider the flat $w_0w_a$CDM model in~\cite{DESI:2024mwx}.} 
Depending on the data set of use (including Pantheon$+$~\cite{Brout:2022vxf}, Union3~\cite{Rubin:2023ovl}, DES-SN5YR~\cite{DES:2024tys} and CMB~\cite{Planck:2018vyg}), the discrepancy can be above $3\s$ level. I call this DESI anomaly in this paper. 
Although it is too early to definitively go beyond the $\Lambda$CDM, I believe that this may be an opportune time to shed light on some approaches for solving the cosmological constant (CC) problem.

One of the enduring theoretical challenges in particle physics and cosmology is the fine-tuning of the CC \cite{Weinberg:1987dv, Weinberg:1988cp}, which is measured as 
\begin{equation} \Lambda_{C} =  \O(10^{-3}) \, \text{eV}.\end{equation}
The CC problem should be addressed by IR dynamics because even the QCD contribution $(1 \, \text{GeV})^4 / (16 \pi^2)$ to the CC must be somehow neutralized, resulting in a tuning of the order $\mathcal{O}(10^{-45})$. Thus solving the CC problem never works at a cosmic temperature or renormalization scale higher than the heavy particle energy scales in the Standard Model of the particle theory because otherwise the radiative corrections would spoil it.

There is a no-go theorem by S. Weinberg stating that a scalar field intended to stabilize the CC and make it vanishingly small in a stationary manner does not work~\cite{Weinberg:1988cp}.\footnote{Exact scale invariance requires fine-tuning of the dimensionless parameter, meaning that a symmetry does not work for the purpose.}
Thus, to solve the CC problem, we are led to consider a CC that varies over time during a cold period of the Universe.
At this point in my reasoning, one may already appreciate the significance of detecting time-varying dark energy, which could be relevant to solving the CC problem.

Concrete models for solving the CC problem with a time-varying CC were proposed by \cite{Banks:1984tw} (see also \cite{Abbott:1984qf}), where a slowly varying scalar field can steer the Universe from expansion to contraction. 
Then the volume of the Universe is maximized at the critical point between the expansion and contraction, at which the CC is vanishingly small. 
Although such a scenario typically predicts an empty Universe, conflicting with Big Bang cosmology, the authors in Refs. \cite{Graham:2019bfu, Ji:2021mvg} showed that the Universe can be reheated by the scalar field in the contracting phase and that the generated plasma induces a bounce, leading to the Big Bang cosmology.
It is interesting to further investigate whether almost scale-invariant primordial density perturbations over the cosmic microwave background (CMB) scales can be generated.

An alternative scenario for the relaxation of the CC during inflation, which is also a period when the Universe is cold if the Gibbons Hawking radiation temperature~\cite{Gibbons:1977mu} is small enough, was proposed by the present author in Ref.\,\cite{Yin:2021uus}. 
The fundamental concept is outlined as follows. Initially, we postulate that the inflationary potential is situated near a critical point where, if the potential's minimum, i.e., the CC, is positive, eternal inflation would ensue. Conversely, if it is negative, the duration of inflation would be limited. 
Then, if there is a scalar field scanning the CC during inflation, the prolonged period of inflation ends when the CC is around zero
After inflation, the CC remains almost constant in the much shorter period of the radiation dominated era than the era for the scanning of the CC. 
Interestingly, in this scenario, the CC in our Universe is naturally non-vanishing and is determined probabilistically, as the end of long inflation in a certain place is also a probabilistic event. Thus given the potential shape of the inflaton, we have prediction of the distribution of the CC relevant to the slope of the scalar potential. 
Although this mechanism requires that the shape of the inflaton potential around such criticality be as finely tuned as the CC, the potential can be rendered stable against radiative corrections, distinguishing it from the original CC tuning.
 The remaining study will be to see if such an inflaton potential can be realized in some UV completion. 

As one can see, neither attempt at explaining the CC is completely successful, but the simplest model that realizes any of the previous setups includes a scalar field, $\f$, with a very flat potential, $V[\f]$, to scan the CC.
This implies that, in the present Universe, we can expand the potential to obtain a linear approximation\footnote{One cannot consider $\f$ has a strong higher order interaction because the no-go theorem by Winberg forbid extrema around the field point. }
\beq
V_C[\f]\simeq V_\f \f,
\eeq 
as the leading-order approximation. Here, we choose the origin of the field $\f$ such that the CC is zero and $V_\f$ is a positive constant without loss of generality.
This potential is defined at a renormalization scale below that of any massive particles in the Standard Model of particle theory, so that $V_\f \f$ represents the dark energy.

In this paper, we show that $\f$ in the linear potential can explain the DESI anomaly consistent with the various setups for the CC problem. In particular, the inflationary CC relaxation scenario  predicts the time-varying dark energy in the way favored by the DESI data. Therefore, the DESI anomaly may provide insights into why the CC is so small. 

Very recently, the authors in Ref.~\cite{Terada} studied the quintessential interpretation of the evolving dark energy (see also past works~\cite{Linder:2002et,dePutter:2008wt}). In particular, the author studied the use of a cosine potential to explain the reported data. 
In my paper, I show that the linear potential of the scalar works to explain the anomaly,  and thus, the DESI anomaly may link to the long-standing cosmological constant problem. 
I also note that the original quintessence models do not solve the CC problem, as the small CC or the absolute potential hight is set by hand.

\section{Big Bang Cosmology with a Linear Potential Scalar Field}

Let us discuss the Big Bang cosmology with homogenous $\f$. The discussion here holds in either \cite{Graham:2019bfu} or \cite{Yin:2021uus}, because the relaxation of the CC happens before the reheating for the current Universe.\footnote{I assumed a low scale slow-roll inflation before the reheating for the primordial density perturbation. This is after the relaxation. It is low-scale and thus the $\f$ fluctuation is negligible. This is guaranteed at least in \cite{Yin:2021uus} for evading the eternal inflation. } 
The cosmological history is as follows: after reheating, we have a radiation-dominated Universe with such a large Hubble parameter $H$ that $\f$ is almost frozen due to friction. 
Then, $H$ decreases as $H \propto a^{-2}$, where $a$ is the scale factor, due to the expansion of the Universe.
In other words, it slow-rolls with a very tiny field excursion, and the field value as well as $V_C$ is kept. 
We can use $V_C = V_{C,\rm ini}$ and $\f = \f_{\rm ini}$ for the initial values of $\f$ and dark energy, respectively, which will be related to the dynamics of the relaxation of the CC. Then,
dark matter becomes the dominant component, and still, $\f$ remains frozen. 
Then $H \propto a^{-3/2}$. As $H$ decreases, the field excursion of $\f$ increases. Therefore, at late times, the change in the potential $V_C$ cannot be neglected if $V_\f$ is not very small. Thus at the early stage we have essentially the same as the $\L$CDM but later we have the time-varying dark energy.

\begin{figure}[!t]
\begin{center}  
   \includegraphics[width=145mm]{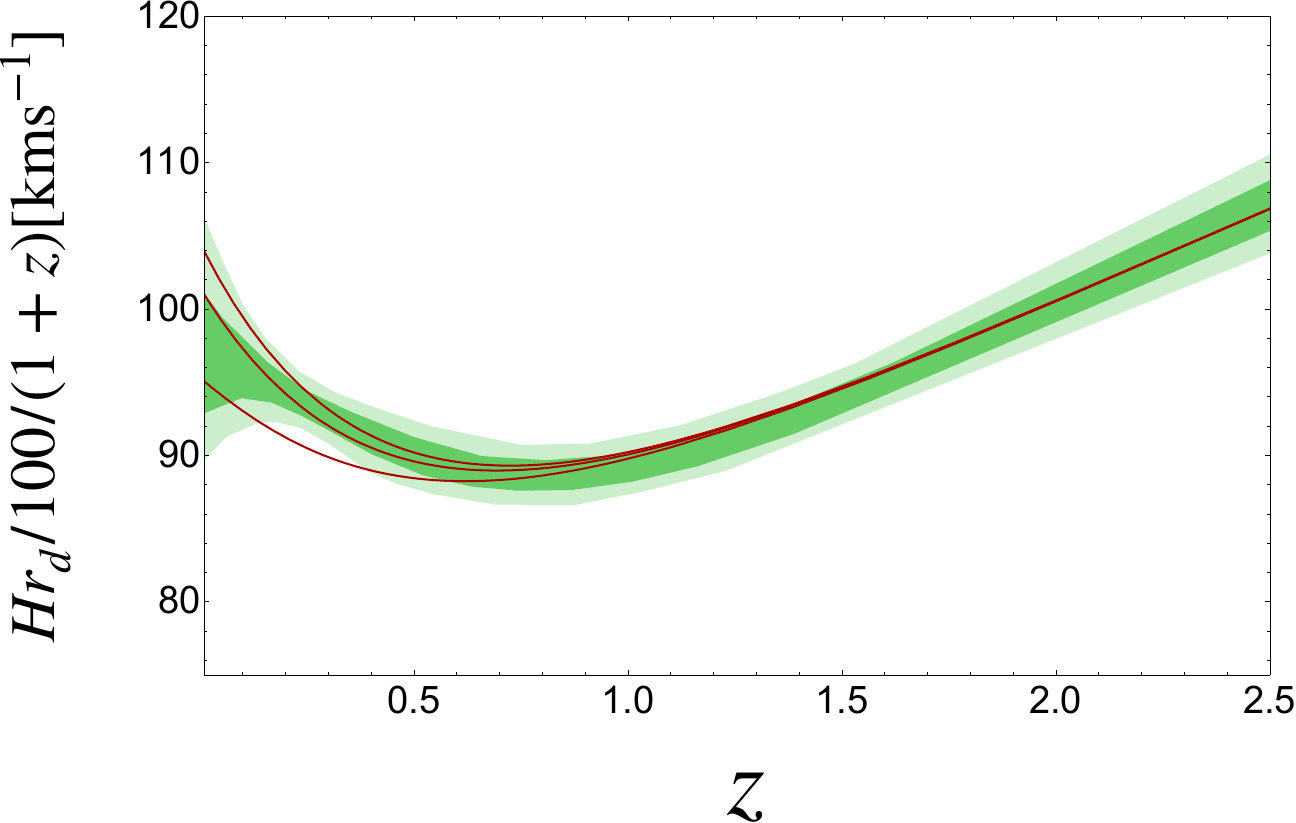}
      \end{center}
\caption{
Expansion history of the late time Universe with a linear potential scalar field. 
The red curves denote the case $V_\f\approx \{3.1,7.6,12\}\times 10^{-66}\GEV^3$ from top to bottom. The green 1 and 2$\s$ regions are taken from \cite{DESI:2024mwx}. We fix $V_{C,\rm ini}\approx 3.0\times 10^{-47}\GEV^4$ and $\rho_m\approx 1.1\times 10^{-47}\GEV^4$. }
\label{fig:1} 
\end{figure}

To see this I solve the equation of motion of the homogeneous part of $\f$ around the present Universe, 
\beq 
\ddot \f +3 H \dot \f =-V_\f,
\eeq  
where $H$ is the Hubble parameter given by 
\beq
H\approx \sqrt{\frac{V_C+\dot{\f}^2/2+\rho_{m}}{3M_{\rm pl}^2}}
 \eeq 
 with $\rho_m$ being the matter density scaling as $\r_m\propto a^{-3}$, and I neglect the other component since I focus on the period dominated by matter/dark energy, around the redshift $z\sim\O(0.1)$. $\dot{X} \equiv dX/dt$, with $t$ being the cosmic time.
$M_{\rm pl} = 2.4 \times 10^{18} \, \text{GeV}$ is the reduced Planck mass. 
  The redshift is related to the Hubble parameter and cosmic time via 
\beq 
\dot{z}[t]  =-(1+z[t])H.
\eeq 
By using $\rho_m \approx 1.1 \times 10^{-47} \, \text{GeV}^4$ (this is $\Omega_m \times 3 H_0^2 M_{\rm pl}^2$, using the fit with DESI+CMB+DESY5 for $w_0w_a$CDM in \cite{DESI:2024mwx}), and $V_{C,\rm ini} = 3.0 \times 10^{-47} \, \text{GeV}^4$, we evaluate the expansion history of the Universe at low $z$.\footnote{I also used several different sets of $\rho_m$ and $H_0$ and find that we can have the prediction within the $1\s$ band in Fig.\,\ref{fig:1}.} This is represented by the red solid lines in Fig.~\ref{fig:1}, where we plot $\frac{H r_d}{100(1+z)}$ as a function of $z$, following Ref.\,\cite{DESI:2024mwx}, with $V_\f = \{3.1, 7.6, 12\} \times 10^{-66} , \GEV^3$ from top to bottom. Here $r_d\approx 147.1$ Mpc is the sound horizon. 
Also shown are the 68\% and 95\% credible regions that fit all the DESI data. One can see that the prediction agrees well with the data.

This parameter choice is consistent with the simplest scenario in \cite{Graham:2019bfu} and is close to the largest possible $|V_\f|$ without causing the Universe to contract before today.
Alternatively,  the choice is close to the  {\it predicted} parameter region of  $\{V_{C,\rm ini}, V_\f\}$  in the scenario in \cite{Yin:2021uus}, as I will demonstrate shortly.

\section{Prediction of Time Varying Dark Energy from CC Relaxation and DESI anomaly}

Now I will discuss the relaxation of the CC in a bit more detail, focusing on the inflationary relaxation by reviewing \cite{Yin:2021uus}.
During inflation, the volume of the Universe, $L_{\rm inf}^3$, that undergoes the `eternal' inflation generically satisfies 
\begin{equation}
\laq{PR}
L^3_{\rm inf}\propto P \cdot R^3\propto e^{\int{(3H_{\rm inf}-P_{\rm esc} )dt}},
\end{equation}
where $H_{\rm inf}$ is the inflationary Hubble parameter satisfying 
$H_{\rm inf}=\sqrt{\frac{V_{\rm inf}+V_C}{3M_{\rm pl}^2}}$. 
Here $V_{\rm inf}[\F]$ is the potential of the inflaton, $\F$, and it is defined to be zero at the minimum of the potential.  
 $P_{\rm esc}$ is the escaping rate of inflation at each Hubble patch. $P_{\rm esc}$ depends on the action for the inflaton. 
For concreteness, we consider a quadratic hilltop inflation model, and then $P_{\rm esc} = C \frac{|V_{\rm inf}''|}{3H_{\rm inf}}$, with $C$ being of $\O(1)$. 
If $3H_{\rm inf} - P_{\rm esc}$ is always positive, `eternal' inflation takes place, and the total volume increases exponentially. If it is negative, the volume decreases. In any case, we have volumes escaping from the regime of `eternal' inflation. 
In this volume, the inflaton $\F$ further undergoes slow-roll inflation to explain the CMB data. This is the case for various successful hilltop inflation models~\cite{Nakayama:2012dw, Guth:2018hsa, Matsui:2020wfx}\cite{Czerny:2014wza, Czerny:2014xja,Czerny:2014qqa,Higaki:2014sja, Croon:2014dma,Higaki:2015kta, Higaki:2016ydn},\cite{Daido:2017wwb, Daido:2017tbr, Takahashi:2019qmh}\cite{Takahashi:2021tff,Narita:2023naj}.

Given that $V_{\rm inf}''$ does not change over time, 
 whether $L_{\rm inf}^3$ increase or decrease depends on the value $H_{\rm inf}$ and thus $V_C$. 
Since $V_C$ is a decreasing function in time (neglecting the quantum diffusion of $\f$), we find that 
the most of volume has $V_C$ around the value that the condition 
\beq
3H_{\rm inf}=P_{\rm esc} 
\eeq
is satisfied. 
By expanding in the leading order of $V_C$ and performing the time integral and considering the slow-roll of $\f$,  we can  derive the distribution of $V_C$ soon after the inflation as 
\begin{equation}
\boxed{ \frac{d}{d V_{C}} L_{\rm end}^3\propto e^{- \frac{ (V_C-V_C^{(0)})^2}{2 \s^2}}, {\rm with~} \s\simeq \frac{V_\f M_{\rm pl}}{\sqrt{6}}}.\end{equation} 
Therefore we see that $V_C$ has typical value around $V_C^{(0)}$, which is determined by the inflation potential $V_{\rm inf}.$ The variance $\s$ is determined by the potential of scanning field $\f$ and also inflaton $\F$. In the analysis, we need to take into account the quantum diffusion of $\f$, but I do not discuss it in this paper for the sake of readability. 
The constraints for neglecting the diffusion effect, as well as other possible constraints such as evading eternal inflation, can be found both analytically and numerically by solving the Fokker-Planck equation in Ref.\,\cite{Yin:2021uus}.

Now, we assume that we have a special inflaton potential, $V_{\rm inf}$, such that $V_C^{(0)}\approx 0$. This is the case where $3H_{\rm inf} = P_{\rm esc}$ when $V_C = 0$, i.e., the inflaton potential alone is at the critical point between eternal inflation and non-eternal inflation. 
This is nothing but a fine-tuning of the potential shape of inflaton and the amount of tuning is as significant as the original CC tuning. However, the shape of the finely-tuned inflaton potential can be made technically natural by symmetry. For instance, we can consider an axion as the inflaton, and the radiative correction to the potential is highly suppressed due to the approximate shift symmetry and exact discrete shift symmetry. The tuning in this sense is different from the one for the original CC problem in which the small CC never becomes technically natural. The important future task is to find a UV completion, which is beyond the scope of this paper. 

The interesting point is that once we admit that $V_{C}^{(0)}$ is small enough we have $V_C$ soon after the inflation,
\beq
V_{C,\rm ini}\sim \s \simeq \frac{V_\f M_{\rm pl}}{\sqrt{6}}\approx 10^{-47}\GEV^4\frac{V_{\f}}{ 10^{-65}\GEV^3}
\eeq
which is close to the value for the good fit in Fig.\ref{fig:1}. 
Although there is a slight tension ($V_{C,\rm ini}\approx 3\times 10^{-47}\GEV^4 $ vs $10^{-47}\GEV^4$), this coincidence seems to be non-trivial. Whether the tension really exists may require further study by performing a data fit not for the $w_0w_a$CDM but  for this model. 
In addition, this tension may be filled by considering that $\f$ couples to the neutrinos, dark matter or other dark radiation inducing a further friction for $\f$~\cite{Berera:1995ie,Berera:1998gx,Yokoyama:1998ju, Graham:2019bfu, Nakayama:2021avl}. 
This allows larger $V_{\f}$ and thus $V_{C,\rm ini}$ for explainning the DESI anomaly, because the motion of $\f$ is further suppressed. The friction is not important during inflation since the particles are absent there. 
In this scenario, $\f$ can mediate force and can be probed phenomenologically~\cite{Moody:1984ba, Pospelov:1997uv}, especially if it couples to usual matter's spin and dark matter~\cite{Kim:2021eye}. If it couples to neutrinos it may induce the cosmic neutrino background (C$\nu$B) decays and may be probed by searching for C$\nu$B.

\section{Conclusions and discussion}

I studied the cosmological evolution with a weakly coupled, very light scalar field with a linear potential, motivated by various proposals to relax the cosmological constant.
I showed that this setup can explain the recently reported preference of time-varying dark energy by DESI. 
In particular, the time-varying dark energy close to the DESI preference is predicted in the inflationary relaxation scenario for the cosmological constant. 

So far I did not specify the origin of the scalar field. 
The scalar field is very likely to have a shift symmetry for the flatness of the potential. This shift symmetry may allow  derivative couplings of the scalar field to the standard model particles or dark matter. 
The coupling to a pair of photons may induce cosmic birefringence (especially there is another anomaly, e.g., \cite{Minami:2020odp}), while couplings to matter may induce spin precessions, and couplings to neutrinos may induce C$\nu$B decays.
Those effects, if observed, can be the smoking gun signal of our scenario. 

Again, I emphasize that relaxation scenarios for the cosmological constant problem seem to have some unnatural aspects. 
Given the new results from DESI, this paper aims to assert that it may be more important to further investigate these directions

\section*{Acknowledgments}
WY was supported by JSPS KAKENHI Grant Numbers 20H05851, 21K20364, 22K14029, and 22H01215.


\begin{thebibliography}{99}
\bibitem{Yin:2021uus}
W.~Yin,
Phys. Rev. D \textbf{106}, no.5, 055014 (2022)
[arXiv:2108.04246 [hep-ph]].

\bibitem{DESI:2024mwx}
A.~G.~Adame \textit{et al.} [DESI],
[arXiv:2404.03002 [astro-ph.CO]].

\bibitem{Brout:2022vxf}
D.~Brout, D.~Scolnic, B.~Popovic, A.~G.~Riess, J.~Zuntz, R.~Kessler, A.~Carr, T.~M.~Davis, S.~Hinton and D.~Jones, \textit{et al.}
Astrophys. J. \textbf{938}, no.2, 110 (2022)
[arXiv:2202.04077 [astro-ph.CO]].

\bibitem{Rubin:2023ovl}
D.~Rubin, G.~Aldering, M.~Betoule, A.~Fruchter, X.~Huang, A.~G.~Kim, C.~Lidman, E.~Linder, S.~Perlmutter and P.~Ruiz-Lapuente, \textit{et al.}
[arXiv:2311.12098 [astro-ph.CO]].

\bibitem{DES:2024tys}
T.~M.~C.~Abbott \textit{et al.} [DES],
[arXiv:2401.02929 [astro-ph.CO]].

\bibitem{Planck:2018vyg}
N.~Aghanim \textit{et al.} [Planck],
Astron. Astrophys. \textbf{641}, A6 (2020)
[erratum: Astron. Astrophys. \textbf{652}, C4 (2021)]
[arXiv:1807.06209 [astro-ph.CO]].

\bibitem{Weinberg:1987dv}
S.~Weinberg,
Phys. Rev. Lett. \textbf{59}, 2607 (1987)

\bibitem{Weinberg:1988cp}
S.~Weinberg,
Rev. Mod. Phys. \textbf{61}, 1-23 (1989)

\bibitem{Banks:1984tw}
T.~Banks,
Phys. Rev. Lett. \textbf{52}, 1461-1463 (1984)

\bibitem{Abbott:1984qf}
L.~F.~Abbott,
Phys. Lett. B \textbf{150}, 427-430 (1985)

\bibitem{Graham:2019bfu}
P.~W.~Graham, D.~E.~Kaplan and S.~Rajendran,
Phys. Rev. D \textbf{100}, no.1, 015048 (2019)
[arXiv:1902.06793 [hep-ph]].

\bibitem{Ji:2021mvg}
L.~Ji, D.~E.~Kaplan, S.~Rajendran and E.~H.~Tanin,
Phys. Rev. D \textbf{105}, no.1, 015025 (2022)
[arXiv:2109.05285 [hep-ph]].

\bibitem{Gibbons:1977mu}
G.~W.~Gibbons and S.~W.~Hawking,
Phys. Rev. D \textbf{15}, 2738-2751 (1977)

\bibitem{Terada}
Y.~Tada, T.~Terada
[arXiv:2404.05722 [astro-ph.CO]].
\bibitem{Linder:2002et}
E.~V.~Linder,
Phys. Rev. Lett. \textbf{90}, 091301 (2003)
[arXiv:astro-ph/0208512 [astro-ph]].

\bibitem{dePutter:2008wt}
R.~de Putter and E.~V.~Linder,
JCAP \textbf{10}, 042 (2008)
[arXiv:0808.0189 [astro-ph]].

\bibitem{Nakayama:2012dw}
K.~Nakayama and F.~Takahashi,
JCAP \textbf{05}, 035 (2012)
[arXiv:1203.0323 [hep-ph]].

\bibitem{Guth:2018hsa}
F.~Takahashi, W.~Yin and A.~H.~Guth,
Phys. Rev. D \textbf{98}, no.1, 015042 (2018)
[arXiv:1805.08763 [hep-ph]].

\bibitem{Matsui:2020wfx}
H.~Matsui, F.~Takahashi and W.~Yin,
JHEP \textbf{05}, 154 (2020)
[arXiv:2001.04464 [hep-ph]].

\bibitem{Czerny:2014wza}
M.~Czerny and F.~Takahashi,
Phys. Lett. B \textbf{733}, 241-246 (2014)
[arXiv:1401.5212 [hep-ph]].

\bibitem{Czerny:2014xja}
M.~Czerny, T.~Higaki and F.~Takahashi,
JHEP \textbf{05}, 144 (2014)
[arXiv:1403.0410 [hep-ph]].

\bibitem{Czerny:2014qqa}
M.~Czerny, T.~Higaki and F.~Takahashi,
Phys. Lett. B \textbf{734}, 167-172 (2014)
[arXiv:1403.5883 [hep-ph]].

\bibitem{Higaki:2014sja}
T.~Higaki, T.~Kobayashi, O.~Seto and Y.~Yamaguchi,
JCAP \textbf{10}, 025 (2014)
[arXiv:1405.0775 [hep-ph]].

\bibitem{Croon:2014dma}
D.~Croon and V.~Sanz,
JCAP \textbf{02}, 008 (2015)
[arXiv:1411.7809 [hep-ph]].

\bibitem{Higaki:2015kta}
T.~Higaki and F.~Takahashi,
JHEP \textbf{03}, 129 (2015)
[arXiv:1501.02354 [hep-ph]].

\bibitem{Higaki:2016ydn}
T.~Higaki and Y.~Tatsuta,
JCAP \textbf{07}, 011 (2017)
[arXiv:1611.00808 [hep-th]].

\bibitem{Daido:2017wwb}
R.~Daido, F.~Takahashi and W.~Yin,
JCAP \textbf{05}, 044 (2017)
[arXiv:1702.03284 [hep-ph]].

\bibitem{Daido:2017tbr}
R.~Daido, F.~Takahashi and W.~Yin,
JHEP \textbf{02}, 104 (2018)
[arXiv:1710.11107 [hep-ph]].

\bibitem{Takahashi:2019qmh}
F.~Takahashi and W.~Yin,
JHEP \textbf{07}, 095 (2019)
[arXiv:1903.00462 [hep-ph]].

\bibitem{Takahashi:2021tff}
F.~Takahashi and W.~Yin,
JCAP \textbf{10}, 057 (2021)
[arXiv:2105.10493 [hep-ph]].

\bibitem{Narita:2023naj}
Y.~Narita, F.~Takahashi and W.~Yin,
JCAP \textbf{12}, 039 (2023)
[arXiv:2308.12154 [hep-ph]].

\bibitem{Berera:1995ie}
A.~Berera,
Phys. Rev. Lett. \textbf{75}, 3218-3221 (1995)
[arXiv:astro-ph/9509049 [astro-ph]].

\bibitem{Berera:1998gx}
A.~Berera, M.~Gleiser and R.~O.~Ramos,
Phys. Rev. D \textbf{58}, 123508 (1998)
[arXiv:hep-ph/9803394 [hep-ph]].

\bibitem{Yokoyama:1998ju}
J.~Yokoyama and A.~D.~Linde,
Phys. Rev. D \textbf{60}, 083509 (1999)
[arXiv:hep-ph/9809409 [hep-ph]].

\bibitem{Nakayama:2021avl}
K.~Nakayama and W.~Yin,
JHEP \textbf{10}, 026 (2021)
[arXiv:2105.14549 [hep-ph]].

\bibitem{Moody:1984ba}
J.~E.~Moody and F.~Wilczek,
Phys. Rev. D \textbf{30}, 130 (1984)

\bibitem{Pospelov:1997uv}
M.~Pospelov,
Phys. Rev. D \textbf{58}, 097703 (1998)
[arXiv:hep-ph/9707431 [hep-ph]].

\bibitem{Kim:2021eye}
D.~Kim, Y.~Kim, Y.~K.~Semertzidis, Y.~C.~Shin and W.~Yin,
Phys. Rev. D \textbf{104}, no.9, 095010 (2021)
[arXiv:2105.03422 [hep-ph]].

\bibitem{Minami:2020odp}
Y.~Minami and E.~Komatsu,
Phys. Rev. Lett. \textbf{125}, no.22, 221301 (2020)
[arXiv:2011.11254 [astro-ph.CO]].
\end{thebibliography}
\end{document}